\documentclass[12pt,oneside,final,a4paper]{article}

\title{Bell's inequality for conditional probabilities as a test for quantum-like behaviour of mind}
\author{Andrei Khrennikov\\
International Center for Mathematical Modeling\\
in Physics and Cognitive Sciences,\\
MSI, University of V\"axj\"o, S-35195, Sweden\\
Email: Andrei.Khrennikov@msi.vxu.se}
\begin{document}
\maketitle

\begin{abstract} We define quantum-like probabilistic behaviour as behaviour which is
impossible to describe by using the classical probability model. We discuss the conjecture that 
cognitive behaviour is quantum-like. There is presented the scheme for an experimental  test 
for quantum-like cognitive behaviour based on a generalization of the famous Bell's inequality.
This generalization is an analogue of Bell's inequality, but for conditional probabilities. 
The use of conditional probabilities (instead of simultaneous probability distributions for pairs
of observables in the original Bell's inequality) gives the possibility to separate two problems 
which are mixed in the original Bell's framework: nonlocality and nonclassical (quantum-like) 
probabilistic behaviour. Our inequality for conditional probabilities can be used for experiments 
with a single system (so we need not to prepare pairs of correlated systems) to find quantum-like
behaviour. This possibility is extremely important in cognitive sciences where it is practically 
impossible to prepare pairs of precisely correlated cognitive systems. Thus the test of the original
Bell's inequality in cognitive and social sciences, or psychology is the problem of huge complexity.
On the other hand, our test based on an analogue of Bell's inequality for 
conditional probabilities can be easily performed in, e.g., psychological experiments or cognitive tasks.

\end{abstract}

\section{Introduction}

It seems that the philosophic system of Whitehead [1]-[3]
was the first attempt to establish a quantum--mental (or more precisely mental $\to$ quantum)
connection. In Whitehead's philosophy of the organism ``quantum" was some feature of basic
protomental elements of reality [1], especially p. 401-403.
See also A. Shimony [4] for modern reconsideration of quantum counterpart of Whitehead's philosophy 
of organism. Unfortunately, the main stream of ``quantum-mental'' investigations was directed 
to {\it quantum reductionism of mental processes,} see [5]-[19]. I do not want to criticize attempts
to perform quantum-reduction of mental processes -- reduction of mind to quantum physical processes.
It is clear that this is the great program of investigations and it is too early to say anything.
I just want to pay attention that, e.g.,  Whitehead did not have in mind  quantum physical reduction 
of mental processes when he wrote about quantum protomental elements of reality. It is  also important 
to mention Bohr's ideas on complementarity in physics, psychology and other domains of science [20] 
(see [21] for detailed 
analysis). We recall the well known fact that, in fact, {\it N. Bohr ``borrowed'' the principle of 
complemetarity from psychology.} In particular, reading of {\it ``Principles of Psychology''}
of W. James  was very important 
for N. Bohr. But N. Bohr neither had in mind quantum-reductionism. He just wanted to emphasize that
some fundamental principles of quantum mechanics  can be valid outside of physics. The crucial for N. Bohr was 
the special information structure of measurement processes in some domains of science which can
differ essentially from measurement processes in classical physics. We also pay attention to
{\it a correspondence between W. Pauli and C. G. Jung.} They neither come to quantum-reductionist
conclusions. We recall a practically forgotten paper of V. Orlov [22]\footnote{which was written
in a concentration camp in 60th and published a few years later.} who proposed to use quantum logic
to describe functioning of brain. Orlov was not looking for reduction of mental processes to quantum physical 
processes. He considered brain as an information system which can in principle use rules of quantum logic.
And, finally, we mention investigations of Bohm-Hiley-Pylkk\"anen [23]--[25]  who proposed to apply the formalism of 
quantum pilot wave (Bohmian
mechanics) to mental processes. They did not try to reduce mental processes to pilot waves coupled
to quantum particles composing the brain. BHP also considered brain as a special information system
which can be described by Bohmian mechanics (see [26], [27] for a rather risky application of BHP-ideas). 

I also think that the most profitable applications of quantum methods to cognitive and social sciences
and psychology can be obtained not via quantum physical reduction, but by considering cognitive systems
as a complex information systems such that some laws of quantum theory can be applied to these
systems (at least with some modifications and generalizations). At this point it is extremely important 
to understand: What is quantum mechanics about?, see [28]. What kind of laws of nature are encoded
in quantum mathematical formalism?  Despite one hundred years of development and great applications 
of quantum theory, there are still no commonly accepted  answers to these questions. 
Quantum theory is characterized by a huge diversity of interpretations, see [29]. My experience as an
organizer of a series of large conferences on foundations of quantum mechanics says that practically 
everybody elaborated its own specific interpretation of quantum mechanics. Even if people use the 
same name, e.g., {\it Copenhagen interpretation,} in reality they can use extremely different versions.
I think that this diversity of interpretations is a sign of a deep crises in foundations 
of quantum mechanics. 

Personally I use so called {\it ensemble interpretation of quantum mechanics} by which quantum mechanics
gives us laws of transformation of probability distributions for ensembles of physical systems. A. Einstein
was one of the strongest supporters of this interpretation, see also Margenau [30], Ballentine [31]. If we use 
this interpretation then at the first sight quantum mechanics does not differ from classical statistical
physics which also gives laws of transformation of probability distributions for ensembles of physical systems.
The crucial point is that laws of transformation of probability distributions are different!  The main exhibition 
of quantum probabilistic behaviour is the {\it interference of probabilities} of alternatives 
which can be easily obtained in experiments with quantum systems, e.g., in the well known
two slit experiment. This interference gives the experimental  basis of the {\it superposition principle.}
The latter is one of the basic elements of a philosophic system of views to the physical reality which 
we can call quantum mysticism. Roughly speaking, since an electron can be in a superposition of states, but
a car cannot, there is a huge gap between domains of application of quantum and classical probability.
There is a rather common opinion that quantum randomness differs crucially from classical randomness.
(the first one is ``irreducible'' and the second one is ``reducible'', see, e.g., [32]);
that quantum theory works in microworld and classical in macroworld. 

These views to quantum theory and especially
probability play the fundamental role in the justification of quantum reductionism. Since laws of 
quantum mechanics work only for quantum particles, if we try to apply these laws to mental 
processes then we should reduce these processes to quantum physical processes. Some supporters of quantum reductionism
reply: ``Not everything is so trivial. We do not claim that there is no macroscopic physical structures
having quantum behaviour; consider, for example, Bose-Einstein condensate.'' This is an interesting argument, but I cannot take
its seriously. Bose-Einstein condensate is a collective effect of quantum systems induced by special 
``quantum conditions.''  By using this example we just disappoint ourselves and scientists working in 
neurophysiology, cognitive sciences and psychology. In fact, we say to them:``You should look 
for these special quantum conditions in brain!'' (in particular, temperature, space and time scales).
I think that R. Penrose is more fair when he wrote that 
an individual neuron could not be in superposition of two states and therefore we cannot use 
quantum mechanics on neuronal level [6]. And in his quantum-mental investigations R. Penrose
operated on fantastically deep microlevel -- level of quantum gravity. Roughly speaking consciousness
is generated by collapses of mass superpositions. Of course, there is the huge gap between the levels
of quantum cosmology and neurophysiology... Quantum reductionism is totally incompatible with modern 
neurophysiology which is based on the neuronal model.

It seems that ``quantum mysticism'' (and, in particular, mystery of quantum probability)
is one of the main sources of quantum reductionism for mental processes.

Recently the problem of quantum probabilities was essentially clarified [33]-[35] when
the interference of probabilities was obtained by using classical ensemble probabilities (and without 
to appeal to the formalism of Hilbert space). In this series of papers there was developed a calculus
of {\it contextual probabilities} -- probabilities depending on contexts. Contexts are  complexes of  
physical, cognitive, social, psychological conditions. It was observed that laws of transformation
of contextual probabilities coincide (under some conditions) with laws induced by the quantum formalism.
But contextual probability is a classical ensemble probability. The crucial point 
is that context dependence of probabilities can be nontrivial. In this case we cannot use 
``absolute'' context-independent  probability measure ${\bf P}.$ We call nontrivial contextual 
probabilistic behaviour (of physical, cognitive, social systems) {\it quantum-like.}

Contextuality of probabilities has no direct coupling with microworld. Even ensembles of macroscopic 
systems can demonstrate nontrivial contextual probabilistic behaviour. Thus quantum-like behaviour
can be in principle found not only for, e.g., electrons or photons, but also for, e.g., human beings or
some ensembles of neurons.\footnote{We dot speak about a quantum-like 
superposition of states of an individual neuron; in our approach interference of probabilities and superposition
are ensemble features.}

In [36] it was proposed to investigate probabilistic quantum-like behaviour for cognitive
and social systems. As the first step we should really show that cognitive systems may exhibit 
quantum-like behaviour. This can be proved only by using experiments (in cognitive science, psychology,...).
Thus we need experimental tests of quantum-like behaviour. In [36] I proposed  a simple test of quantum-like behaviour
based on interference of probabilities for two questions $a$ and $b$ which can be asked to people. 
In [37] there were performed experiments with students giving preliminary confirmation of quantum-like 
behaviour of people. Up to now there have not been performed experiments with animals, fishes or insects.
Thus there is no experimental data on quantum-like behaviour for these biological organisms. It is natural 
to suppose that only biological organism having a rather high level of mental development can 
exhibit quantum-like behaviour. For example, we can speculate that insects would always behave classically.
We can measure numerically the level of quantum-like behaviour by the magnitude of the coefficient 
interference of probabilities, see [36].

In this paper I propose a new more complex test of quantum-like behaviour for people
(but similar experiments can be also performed for animals) based on three questions (or cognitive tasks) $a, b, c.$
This  test is based on a generalization of the famous Bell's inequality, see, e.g., [38].
This generalization is an analogue of Bell's inequality, but for conditional probabilities. 
It was derived in [39]. The use of conditional probabilities (instead of simultaneous probability distributions for pairs
of observables in the original Bell's inequality) gives the possibility to separate two problems 
which are mixed in the original Bell's framework: nonlocality and nonclassical (quantum-like) 
probabilistic behaviour. J. Bell (who was nonlocal realist) wanted to show that quantum world is nonlocal: 
by performing a measurement on one of correlated particles we change the state of the second particle.
We are not interested in the use of Bell's inequality for
investigation of mental nonlocality.\footnote{We do not 
reject such a possibility , see, e.g., [40] on interesting results in this direction.}
We are very pragmatic. We ``just'' want to show that some  cognitive systems are 
quantum-like. This will give us the possibility to to apply powerful methods of quantum mechanics
to cognitive and social sciences, psychology, economy,....

Our inequality for conditional probabilities can be used for experiments 
with a single cognitive system  to find quantum-like
behaviour (so we need not to prepare pairs of ``correlated cognitive systems''). 
This possibility is extremely important in cognitive sciences where it is practically 
impossible to prepare pairs of precisely correlated cognitive systems. Thus the test of the original
Bell's inequality in cognitive and social sciences, or psychology is the problem of huge complexity.
On the other hand, our test based on an analogue of Bell's inequality for 
conditional probabilities can be easily performed in  psychological experiments or cognitive tasks 
for animals and people.

We hope that this paper will attract attention of scientists working in  experimental cognitive
science, psychology, social sciences and that there will be really found violations of an analogue
of Bell's inequality for conditional probabilities presented in this paper.

\section{Description of test}

The main aim of our paper is to attract attention of experimenters working in cognitive
science, psychology, sociology to our test of quantum-like behaviour. Therefore we would
not like to disturb these people by mathematical derivations and discussions on the analogy
with quantum physics. The test by itself is very simple and we present it here, see 
section 3 for derivations and discussions.

Let $a, b, c$ be three questions which can asked to people (it can also be some 
cognitive tasks for people or animals). The answers ``yes'' and ``no''  are encoded by 
+1 and -1, respectively. 

1). We collected an ensemble $S$ of people which are ``prepared'' in the same 
mental state with respect to these  questions.\footnote{For example, if questions $a, b, c$
about politics then it is assumed that people from $S$ live in the same country, have 
the same social and cultural level, the same age, sex  and so on.} 

The crucial in our test is that an ensemble  $S$ should be homogeneous with respect to questions
$a,b,c$  in the sense that for all questions the probabilities of answers are 
equal:
$$
{\bf P}(u=+1)={\bf P}(v=-1)=\frac{1}{2},\; \mbox{where}\; u=a,b,c.
$$
In our experimental test we consider frequencies instead of probabilities,
${\bf P}(u=+1) \approx \nu(u=+1), u=a,b, c,$ where:
$$
\nu(u=+1)= \frac{\mbox{the  number   of   people  who  answered  ``yes''  to  the  question} \; u}
{\mbox{the  total  number  of  people  in} \; S}
$$

2). We split  $S$ into two subensembles $U$ and $V$ containing  equal numbers of people.

3). The question $b$ is asked to people from the ensemble $U$ and the 
question $c$ is asked to people from the ensemble $V.$ Then we select 
in the ensemble $U$ people who answered  ``yes'' and  people who answered  ``no''.
Corresponding subensembles are denoted by symbols $U_+^b$ and $U_-^b,$ respectively.
We do the same with people from $V,$  and get subensembles $U_+^c$ and $U_-^c.$

4). We ask the question $a$ to people belonging the ensembles $U_+^b, V_+^c$ and 
the question $c$ to people belonging the ensemble $U_-^b.$ By counting frequencies we find 
approximately conditional probabilities; for example, ${\bf P}(a=+1/b=+1)\approx \nu(a=+1/b=+1),$
where 
$$
\nu(a=+1/b=+1)=  \frac{\mbox{the  number  of  people  who  answered  ``yes''  to} \; a}
{\mbox{the  total  number  of  people  in} \; U_+^b\; }
$$

Our text of quantum-like behaviour is based on the violation of the following inequality
(an analogue of Bell's inequality for conditional probabilities), see section 3:
\begin{equation}
\label{BBA}
{\bf P}(a=+1/b=+1) + {\bf P}(c=+1/b=-1) \geq {\bf P}(a=+1/c=+1)
\end{equation}
 
If this inequality is violated than people (or animals) behave in the quantum-like way with respect 
to questions $a,b,c.$

In statistical investigation we should fix small  $\delta>0$ and probability $p\approx 1.$  Calculate
$$
\Delta(a,b,c)= \nu(a=+1/c=+1) - \nu(a=+1/b=+1) - \nu(c=+1/b=-1)
$$
and apply the criteria $\chi^2$ for the hypothesis that $\Delta(a,b,c) \geq \delta.$

\section{Inequality for conditional probabilities}

{\bf 3.1. Quantum-like probabilistic behaviour.} Since the first days of quantum 
mechanics there is intensively discussed the {\it problem of realistic description} --
the possibility to use in quantum mechanics the same realistic viewpoint to observables
as in classical physics: values of physical observables as properties of objects. We recall
that quantum mechanics is a statistical theory and it cannot say anything about behaviour of individual
quantum systems (at least we use such a viewpoint to quantum mechanics in this paper).
It seems that first time the rigorous mathematical definition of statistical realism was given by J. Bell 
[38]. By analogy with quantum theory we give the following definition:

{\bf Definition 1.} {\it Let ${\cal O}=
\{a, b, \ldots c\}$ be a family of observables (physical, or cognitive, or psychological, or social). 
This family permits the realistic statistical description if there 
exists a Kolmogorov probability space ${\cal K}=(\Omega, {\cal F}, {\bf P})$ such that all observables 
belonging to ${\cal O}$ can be represented by random variables on ${\cal K}.$}

Here $\Omega$ is an ensemble of, e.g., physical systems, or neurons, or neural networks,
or animals, or people, or collectives of people,... (in mathematical model $\Omega$ is an ideal 
infinite ensemble); ${\cal F}$ is a $\sigma$-field of subsets of $\Omega$ and ${\bf P}$ is 
a probability measure on ${\cal F}.$ We recall that conditional probabilities in the Kolmogorov model  are defined by 
the Bayes' formula: 
$$
{\bf P}(a=a_1/b=b_1)=\frac{{\bf P}(a=a_1, b=b_1)}{{\bf P}(b=b_1)}.
$$

{\bf Definition 2.} {\it A  family of observables ${\cal O}= \{a, b, \ldots c\}$ 
which does not permit the realistic statistical description is called quantum-like.}

Let ${\cal O}$ is a quantum-like family of observables and observations are performed on 
some class of systems (e.g., physical systems, or neurons, or neural networks,
or animals, or people, or collectives of people,...). We say that such systems exhibit 
{\it quantum-like behaviour.} Our aim is to present an experimental test for quantum-like 
behaviour for three observables, $a,b,c.$ 

In quantum mechanics J. Bell studied similar (but not the same!) problem. He considered the famous
Einstein-Podolsky-Rosen experiment [41].\footnote{In fact, the Bohm-version of the original EPR-experiment.
In the EPR-experiment there are considered continuous variables, position and momentum; in the EPR-Bohm-experiment
dichotomous variables, e.g., spin projections.}
There are prepared pairs of correlated systems $\omega= (s, s^\prime).$ 
Let $a, b, c$ be three dichotomous observables on $s$ and $a^\prime, b^\prime, c^\prime$ corresponding observables
on $s^\prime.$ For example $a, b, c$ are measurements on $s$ of spin projections corresponding to angles $\theta_1, \theta_2, \theta_3$
and $a^\prime, b^\prime, c^\prime$ are spin projections corresponding to same angles $\theta_1, \theta_2, \theta_3,$
but the latter observations are  performed on $s^\prime.$ It is supposed that we have precise anti-correlations:
\begin{equation}
\label{AC}
a(s)= - a^\prime(s^\prime), \; b(s)= - b^\prime(s^\prime), \;  c(s)= - c^\prime(s^\prime).
\end{equation}
Then J. Bell proposed to consider pair vise correlations of variables $u=a,b,c$ and $v=a^\prime, b^\prime, c^\prime$
and starting with assumption of realism and locality he proved inequality for these correlations. This inequality
is violated by quantum correlations. 

We remark that in quantum theory measurements have huge disturbance effect therefore we cannot measure, e.g., $a$ on $s$
without to disturb $s.$ Thus we should use different $s$ for different measurements (for, e.g., $a$ and $b).$ 
For a single system $s$ we cannot perform the simultaneous measurement of $a$ and $b;$ we cannot find the 
simultaneous probability distribution of $a$ and $b$ and their correlation. Therefore
Einstein, Podolsky and Rosen proposed to consider pairs of correlated particles and therefore J. Bell was 
successful with his test for correlations (for observations on correlated particles).

We would like to study observables in cognitive science, psychology, sociology. In these frameworks 
it seems to be impossible to realize preparations of ensembles which would satisfy to
the condition of precise (anti)-correlation. Therefore we need Bell's like tests for measurements
performed on a single system. Of course, such a framework has nothing to do with investigations on
mental nonlocalty. We study only quantum-like behaviour of mental systems.

 {\bf 3.2. Wigner inequality.} We shall use the following simple mathematical result: 

{\bf Theorem 1.} (Wigner inequality) {\it{Let $a, b, c= \pm 1$ be arbitrary
dichotomous random variables on a single Kolmogorov space ${\cal K}.$ Then the following inequality holds true.}}
\begin{equation}
\label{BB}
{\bf P} (a=+1, b=+1) + {\bf P}(b=-1, c=+1) \geq {\bf P}(a=+1, c=+1)
\end{equation}

The proof of this theorem in purely mathematical framework can be found e.g., in my book [42], p. 89-90. However,
the inequality (\ref{BB}) is, in fact, the well Wigner's inequality, see [43]. Wigner proved (\ref{BB}) for 
three random variables, but then he applied (\ref{BB})
to the EPR-Bohm experiment for correlated particles. By using the condition of the precise anticorrelation
(\ref{AC}) he wrote (\ref{BB}) in the form:
\begin{equation}
\label{BBU}
{\bf P} (a=+1, b^\prime=+1) + {\bf P}(b=-1, c^\prime=+1) \geq {\bf P}(a=+1, c^\prime=+1)
\end{equation}
It is easy to see that (\ref{BBU}) is violated for an appropriative choice of spin projectors.

By reasons presented in section 2.1 we would not like to use (\ref{BBU}) as an experimental test.
But we neither can use directly (\ref{BB}). It is evident that, e.g., psychological  observations 
can also have disturbance effect and we cannot find a simultaneous probability distribution 
for some  pairs of questions $a$ and  $b.$

{\bf 2.3. Inequality for conditional probabilities.}
As a simple consequence of Theorem 1 we obtain: 

{\bf Theorem 2.} (``Bell's inequality for conditional probabilities''). {\it{Let $a, b, c=\pm 1$ be 
dichotomous symmetrically distributed random variables on a single Kolmogorov space. Then the following inequality holds true:}}
\begin{equation}
\label{BB1}
{\bf P}(a=+1/b=+1) + {\bf P}(c=+1/b=-1) \geq {\bf P}(a=+1/c=+1)
\end{equation}

{\bf Proof.} We have ${\bf P}(b=+1)={\bf P}(b=-1) = {\bf P}(a=+1)={\bf P}(a=-1)={\bf P}(c=+1)={\bf P}(c=-1)=1/2.$

Thus 
$$
{\bf P}(a=+1/b=+1) + {\bf P}(c=+1/b=-1)=2{\bf P}(a=+1, b=+1) + 2{\bf P}(c=+1, b=-1)
$$
and 
$$
{\bf P}(a=+1/c=+1)=2{\bf P}(a=+1, c=+1).
$$ 
Hence by (\ref{BB}) we get (\ref{BB1}).

As was shown in [39], the inequality (\ref{BB1}) is violated in quantum mechanics (by 1/2-spin
projections).

We underline again that the main distinguishing feature of (\ref{BB1}) is the presence 
of {\bf only conditional probabilities.} Conditional probabilities can always be calculated:
we first ask the question $b$, then select all people who answered ``yes'', and then ask
them the question $a.$

\medskip

{\bf References}

1. A. N. Whitehead, 
{\it Process and Reality: An Essay in Cosmology.}  Macmillan Publishing Company, New York(1929).

2. A. N. Whitehead,  {\it  Adventures of Ideas.}  Cambridge Univ. Press, London (1933)

3. A. N. Whitehead, {\it Science in the modern world.} Penguin, London  (1939).

4. A. Shimony, 
On Mentality, Quantum Mechanics and the Actualization of Potentialities. 
In R. Penrose, M. Longair (Ed.) {\it The large, the small and the human mind.}
 Cambridge Univ. Press, New York(1997).

5. R. Penrose, {\it The emperor's new mind.} Oxford Univ. Press, New-York (1989).

6. R. Penrose  {\it Shadows of the mind.}  Oxford Univ. Press, Oxford (1994).

7. P. M. Churchland, {\it Matter and consciousness.} MIT Press, Cambridge (1999).

8. H. P. Stapp (1993) {\it Mind, matter and quantum mechanics.}  Springer-Verlag, Berlin-New York-Heidelberg.

9. M. Jibu, K. Yasue, A physical picture of Umezawa's quantum brain dynamics.
In {\it Cybernetics and Systems Research}, ed. R. Trappl (World Sc., London, 1992).

10. M. Jibu, K. Yasue, {\it Quantum brain dynamics and consciousness.}
J. Benjamins Publ. Company, Amsterdam/Philadelphia. 

11. H. Umezawa, {\it Advanced field theory: micro, macro, and thermal physics.}
American Inctitute of Physics, New-York (1993).

12. G. Vitiello, {\it My double unveiled - the dissipative quantum model of brain.}
J. Benjamins Publ. Company, Amsterdam/Philadelphia (2001).

13. M. Lockwood, {\it Mind, Brain and Quantum.} Oxford,Blackwell (1989).

14. J. A. Barrett, {The quantum mechanics of minds and worlds.} Oxford Univ. Press, 1999.

15. M. Pitk\"anen, {\it TGD inspired theory of consciousness with applications to biosystems.}
Electronic vesrion: http://www.physics.helsinki.fi/~matpitka.

16. S. Hameroff, Quantum coherence in microtubules. 
A neural basis for emergent consciousness? {\it J. of Consciousness Studies,}
{\bf 1,} 91-118 (1994); Quantum computing in brain microtubules? The Penrose-Hameroff Orch Or model
of consciousness. {\it Phil. Tr. Royal Sc., London,} A, 1-28 (1998).

17. N. E.Mavromatos, D. V. Nanopoulos, I. Samaras, K. Zioutas,
{\it Advances in structural biology}, {\bf 5}, 127-134 (1998).

18. A. Mershin, D. V. Nanopoulos, E. M. C. Skoulakis, Quantum brain? Preprint
http://xxx.lanl.gov/abs/quant-ph/0007088 (2000).

19. A. Priel, N. Wolf, S. Hameroff, Dynamical properties of dendric arrays of microtubules - relevance to 
consciousness. Proc. Int. Conf. {\it Toward a
Science of Consciousness,} p. 69, Tucson, Arizona (2002).

20. N. Bohr, {\it The philosophical writings of Niels
Bohr}, 3 vols., Woodbridge, Conn.: Ox Bow Press (1987).

21. A. Plotnitsky, Reading Bohr:
Complementarity, Epistemology, Entanglement, and Decoherence.
Proc. NATO Workshop ''Decoherence and its Implications for Quantum
Computations'', Eds. A.Gonis and P.Turchi, p.3--37, IOS Press,
Amsterdam (2001).

22. Y. F. Orlov,  The wave logic of consciousness: A hypothesis. 
{\it Int. J. Theor. Phys.} {\bf 21}, 1, 37-53 (1982) . 

23. D. Bohm,  and B. Hiley , {\it The undivided universe:
an ontological interpretation of quantum mechanics.}
Routledge and Kegan Paul, 
London (1993).

24. B. Hiley, P. Pylkk\"anen,  Active information and cognitive science -- 
A reply to Kiesepp\"a. In: Brain, mind and physics.  Editors: Pylkk\"anen, P., Pylkk\"o, P., Hautam\"aki, A..
IOS Press, Amsterdam (1997).

25. B. Hiley, Non-commutavive geometry, the Bohm interpretation and the mind-matter relationship.
To appear in Proc. CASYS 2000, LIege, Belgium, 2000.

26. A. Yu. Khrennikov,  Classical and quantum mechanics on $p$-adic trees of ideas.
{\it BioSystems,}  {\bf 56}, 95-120  (2000)

27. A. Yu. Khrennikov,  Classical and quantum mechanics on information spaces
with applications to cognitive, psychological,
social and anomalous phenomena. {\it  Found. Phys.} {\bf 29,}  1065-1098 (1999).

28. A. Yu. Khrennikov, What is really ``quantum'' in quantum theory? 
{\it Foundations of Probability and Physics-2,} 
Ser. Math. Modelling in Phys., Engin., and Cogn. Sc., vol. 5, 285-296,
V\"axj\"o Univ. Press, 2002.

29. Proc. Int. Conf. {\it Quantum Theory: Reconsideration
of Foundations.} Ed.: A. Yu. Khrennikov, Ser. Math. Modelling in Phys., Engin., and Cogn. Sc., vol.2,
163-196, V\"axj\"o Univ. Press ( 2002).

30.  H. Margenau, {\it Phil. Sci.,} {\bf 25}, 23 (1958).

H. Margenau, {\it Annals of Physics}, {\bf 23}, 469-485 (1963).

31. L. E. Ballentine,  The statistical interpretation of quantum mechanics,
{\it Rev. Mod. Phys.,} {\bf 42}, 358--381 (1970). 

32. A. Zeilinger, On the interpretation and philosophical foundations of
quantum mechanics.
in {\it  Vastakohtien todellisuus.} Festschrift for K.V. Laurikainen.
U. Ketvel et al. (eds), Helsinki Univ., Press (1996).

33. A. Yu. Khrennikov, Origin of quantum probabilities. Proc. Conf.
"Foundations of Probability and Physics", {\it Quantum Probability and White Noise Analysis},
{\bf 13}, 180-200, WSP, Singapore (2001).

34. A. Yu. Khrennikov, Linear representations of probabilistic transformations induced
by context transitions. {\it J. Phys.A: Math. Gen.,} {\bf 34}, 9965-9981 (2001).

35. A. Khrennikov, Ensemble fluctuations and the origin of quantum probabilistic rule.
{\it J. Math. Phys.}, {\bf 43}, N. 2, 789-802 (2002).

36. A. Khrennikov, {\it On cognitive experiments to test quantum-like behaviour 
of mind.} {\it Reports from V\"axj\"o University,} N 7 (2002);
http://xxx.lanl.gov/abs/quant-ph/0205092 (2002).

37. E. Conte, O. Todarello, A. Federici, F. Vitiello, M. Lopane, A. Khrennikov,
A preliminar evidence of quantum-like behaviour in measurements of mental states.
Reports from MSI, N. 03090, 2003. http://xxx.lanl.gov/abs/quant-ph/0307201. 

38. J. S. Bell, {\it Speakable and unspeakable in quantum mechanics.}
Cambridge Univ. Press (1987).

39. A. Yu. Khrennikov, Bell's inequality fo conditional probabilities and its violation by 
one particle quantum system. Report from MSI, Vaxjo Univ., N 03089 (2003).
http://xxx.lanl.gov/abs/quant-ph/0308078.

40. F. Thaheld, Biological nonlocality and the mind-brain interaction problem:
comments on a new empirical approach. {\it Biosystems}, {\bf 70}, 35-41 (2003).

41. A. Einstein, B. Podolsky, N. Rosen,  Phys. Rev., {\bf 47}, 777--780
(1935).

42.  A. Yu. Khrennikov, Interpretations of Probability. VSP Int. Sc. Publishers, Utrecht/Tokyo (1999).

43. E. P. Wigner, On hidden variables and quantum mechanical probabilities. 
{\it Am J. Phys.}, {\bf 38}, 1005 (1970).

\end{document}